\documentclass[a4paper,twoside,10pt]{article}

\usepackage{cite}

   \usepackage[pdftex]{graphicx}
\usepackage[cmex10]{amsmath}

\renewcommand{\vec}[1]{\mathbf{#1}}
\newcommand{\dif}{\mathrm{d}}

\newcommand{\rx}{r_{x}}
\newcommand{\ry}{r_{y}}
\newcommand{\rz}{r_{z}}
\newcommand{\Xx}{\vec{\hat{\imath}}}
\newcommand{\Yy}{\vec{\hat{\jmath}}}
\newcommand{\Zz}{\vec{\hat{\it{k}}}}
\newcommand{\Roff}{R_{\mathrm{off}}}
\newcommand{\Ron}{R_{\mathrm{on}}}

\begin{document}
%
\title{The Memory-Conservation Theory of Memristance}


\author{Ella Gale
\textit{Unconventional Computing Group}\\
Bristol Robotics Laboratory\\
Bristol, UK\\
Email: ella.gale@brl.ac.uk
}


%


\maketitle

\begin{abstract}
The memristor, the recently discovered fundamental circuit element, is of great interest for neuromorphic computing, nonlinear electronics and computer memory. It is usually modelled either using Chua's equations, which lack material device properties, or using Strukov's phenomenological model (or models derived from it), which deviates from Chua's definitions due to the lack of a magnetic flux term. It is shown that by modelling the magnetostatics of the memory-holding ionic current (oxygen vacancies in the Strukov memristor), the memristor's magnetic flux can be identified as the flux arising from the ions. This leads to a novel theory of memristance consisting of two components: 1. A memory function which describes how the memristance, as felt by the ions, affects the conducting electrons located in the `on' part of the device; 2. A conservation function which describes the time-varying resistance in the `off' part of the device. This model allows for a straight-forward incorporation of the ions within 
the electronic theory and relates Chua's constitutive definition of a memristor with device material properties for the first time.
\end{abstract}


%

\section{Introduction}
The memristor is the 4$\mathrm{th}$ fundamental circuit element, predicted~\cite{14}, due to completeness principles, to exist alongside the resistor, capacitor and inductor. It relates charge, $q$, that has flowed through a circuit with magnetic flux, $\varphi$ by the relation
\begin{equation}
d \varphi (t) = M(q) d q(t) \; ,
\label{eqn:Chua}
\end{equation}
where $M$ is the memristance. A memristor is identified by its pinched hysteresis loop in $V$-$I$ space~\cite{14} and frequency dependence~\cite{279}. The memristor is the first non-linear circuit element. It has been suggested for next-generation computer memory~\cite{15}, due to its d.c. response~\cite{243} and memory~\cite{239} it has been suggested for neuromorphic computing applications, and, as its state-carrying components are oxygen vacancies, it has been suggested that it would be more resilient to cosmic rays and could have uses in resilient electronics~\cite{Oli1}. Its low power operation and ability to hold a state suggests it will be useful in green electronics. 

\begin{figure}[!t]
\includegraphics[width=3.5in]{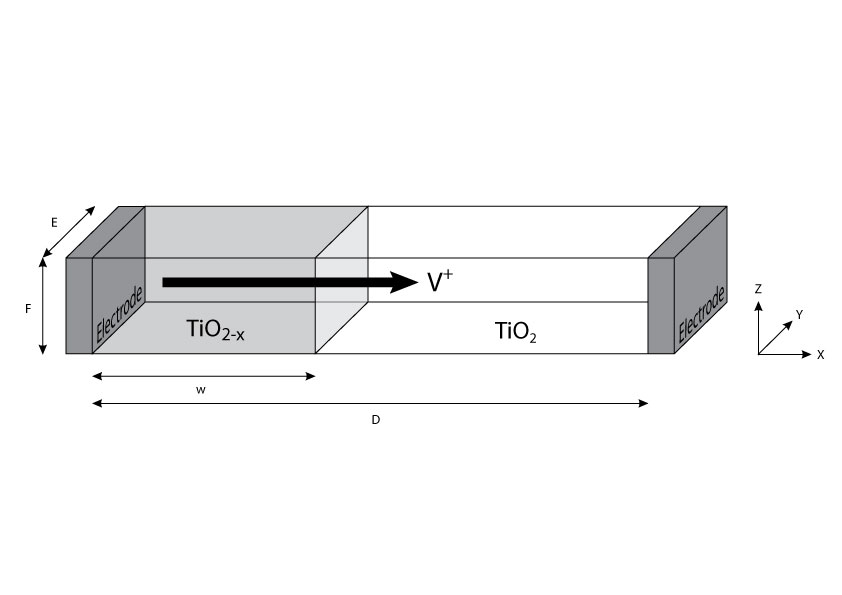}
\caption{\label{fig1} The Strukov memristor. The shaded area is doped low-resistance titanium dioxide, the unshaded area is stoichiometric high-resistance titanium dioxide. Vacancy, V$^{+}$, movement through the material is shown by the arrow. $w$ marks the boundary between the two forms of titanium dioxide.}
\end{figure}

The first instantiation of the device recognised to exist was the Strukov memristor~\cite{15} (although it had been created earlier~\cite{Review1}). The Strukov memristor (also known as the HP memristor after the company that owns the work) is a material with a high resistance part, TiO$_2$ (which has the resistance of $\Roff$), and a low resistivity material, TiO$_{2-x}$ which is doped by oxygen vacancies (which has the resistance $\Ron$), and a boundary between the two materials $w$, which moves as the material inter-converts from one form to another, see figure~\ref{fig1}. ~\cite{15} presented a simple model for the memristor's operation by modeling a uniform field across the ON part of the device (TiO$_{2-x}$), neglecting the effect of the field across the OFF part of the device, and assuming both linear ionic drift through $\Ron$ (which makes it a bulk property~\cite{294}) and instantaneous ohmic conduction. As the theory stands in~\cite{15}, $w$ acts un-physically at the boundaries, this effect is 
corrected and non-linearities are introduced at the device boundaries by the addition of window functions (see for example~\cite{2,46,305}). Most theoretical research has concentrated on using either Chua's theory~\cite{14,84}, Strukov's model~\cite{15} or an extension of it~\cite{224}, although there has been some work in extending the model to a quantum domain~\cite{93}. In this paper, we will derive a novel description of the Strukov memristor by considering the electromagnetics of a uniform field across the whole device. This approach allows us to model the oxygen vacancies separately from the conducting electrons, leading naturally to a two-level model with similarities to the separation of vibrations and electronic transport in molecular electronics~\cite{PhD}, and provides an answer for what the magnetic flux in equation~\ref{eqn:Chua} could be associated with. 

\section{Introducing the Memory Property}

The following analysis is general but will be discussed using the example of the Strukov memristor. The question of whether the Strukov memristor is a true Chua memristor will be approached by first asking how a Chua memristor would behave based on Chua's equations. We are going to focus on the physical property of the memristor responsible for its memory, which we shall call the `memory property'.

For there to be a memory in a memristor, we shall postulate that the memory property must be both separate from the conducting electrodes, and slower to respond to a voltage change than the conducting electrons. This slower response time leads to the lag in current which gives rise to the hysteresis loop and explains the frequency dependence of memristance: if the voltage changes too fast for the memory property, it can't respond fast enough for a measurable change and the size of the hysteresis loop shrinks to a straight line (this is the ohmic regime). 

The memory property has to respond to the voltage, which suggests that it either needs to be affected by the potential difference and therefore be charged, or to undergo a structural change due to the electrical energy supplied. Note that to make a memristor, rather than a memristive system, the device property that causes this change in resistance must be controlled by voltage -- this is necessary for the memristor to be a two-terminal device which is part of Chua's definition for a fundamental circuit element memristor. And, for the memristor to be of any real use, this change in memory property has to be (at least qualitatively) reversible, so the device can switch back and forth. 

To be explicit, the memory property is the physical property or part of the system which holds the system's state, and the state variable is the theoretical label of the aspect of the memory property which is $q$ in equation~\ref{eqn:Chua}. We expect these to be related.

Strukov et al assumed that $q$ (the state variable in equation~\ref{eqn:Chua}) should be the conducting electronic charge, $q_{e}$ (or rather the total charge in the system which is overwhelmingly the electronic charge), as they derive it by integrating the ohmic electronic current over time, but this is impossible because the conducting electrons cannot travel more slowly than themselves and thus the electronic charge can not be a cause of hysteresis. Therefore, another aspect of the system must be responding to the voltage on a different time-scale and it is this response which affects the electronic current. 

The property that best fits the criteria outlined for the memory property is the oxygen vacancies because the state of the memristor is stored by them (as they do not dissipate when the voltage is removed), they drift slower than the conducting electrons which introduces the hysteretic lag (which is recorded in the electronic current because their presence changes the resistivity of the material) and they respond to voltage. If the vacancies are the memory property, then it is their charge, $q_{v}$, which is the state variable which should be used in Chua's equation. Thus, although it is the effect on the electronic current which is measured (and will be of use in real world devices), the electronic current is irrelevant in the actual process of memristance.

Note, Strukov et al were aware that the vacancies are the physical property responsible for the memory, however, they did not relate it to the state variable in Chua's equation, instead assuming that it was the electronic charge, the time integral of electronic current, which was important. 

To derive an alternative model of the memristance of the Strukov memristor, we shall calculate the magnetic flux associated with vacancy motion.

\section{Calculating the Magnetic Flux}

The schematic for the Strukov memristor is shown in figure~\ref{fig1}. Note that the $x$ direction is taken as being the direction of ion flow, with $D$ being the limit of the titanium dioxide layer in the direction (i.e. it's thickness which is 10nm~\cite{15}) and $w(t)$ being the position of the boundary where $0<w(t)<D$. The $y$ and $z$ axes are in the plane of the electrodes with the limits $E$ and $F$ and are both 50nm in the crossbar memristor. We take the terms of our integral in the coordinate system for inside the memristor, i.e.: $r_x$, $r_y$ and $r_z$. To be explicit about our starting assumptions, we assume a linear boundary and that the memory property required from our analysis of Chua's equations is the oxygen ions/oxygen vacancies.

\subsection{Calculating the Magnetic Feild Due to the Oxygen Vacancy Current}

To calculate the flux which should be in Chua's equation, we start by calculating the flux associated with a steady-line current, and this is given by the Biot-Savert law for the magnetic field associated with a volume current. This is the most appropriate formulation of the Biot-Savert law because we are going to consider the magnetic flux just above the memristor surface where the memristor is best viewed as a 3-dimensional object. The Biot-Savert law comes from magnetostatics, a branch of electromagnetic theory that describes the magnetic effects due to constant currents, although our current will change, magnetostatics is still a valid approach because the changing current is, in this case, far slower than that to which such theory is successfully applied (namely mains A.C. (50-60Hz))~\cite{Electrodynamics}.

From this expression and using the Biot-Savert law, the magnetic field (also known as magnetic flux density), $\vec{B}$, at a point, $p$, associated with with this charge is given by the Biot-Savert law for a volume current, $\vec{J}$:

\begin{equation}
\vec{B}(p) = \frac{\mu_0}{4 \pi} \int \frac{J \vec{d\hat{J}} \vec{\times} \vec{\hat{r}}}{r^{2}} \dif \tau
\label{eq:BiotS}
\end{equation}

where $\mu_0$ is the permittivity of a vacuum, $\dif \vec{\hat{J}}$ and $\dif \vec{\hat{r}}$ are the unit vectors for $\vec{J}$ and $\vec{r}$ where $\vec{r}$ is the vector of length $r$ from the volume infinitesimal $\dif \tau$ to point $p$, given by $\vec{r} = \{ r_{x} \vec{\hat{i}}, r_{y} \vec{\hat{j}}, r_{z} \vec{\hat{k}} \}$. 

The magnetic field integral in equation~\ref{BTripInt} is taken over the volume of the device that contains flowing vacancies (which is $w \times E \times F$). This volume is time-dependent due to $w$, but at an instant in time, $t$, the magnetic field is given by

\begin{equation}
 \vec{B}(p,t) = \frac{\mu_0}{4 \pi} \int_{0}^{F} \int_{0}^{E} \int_{0}^{w(t)} \frac{ \vec{J} \vec{\times} \vec{r} }
{|\vec{r}|^{3}} \dif r_{x} \dif r_{y} \dif r_{z} \:,
\label{BTripInt}
\end{equation}

where we have expanded the volume integral to a 3-D Cartesian space and $|\vec{r}|^{3}$ is the cube of the length of vector $\vec{r}$, and the change in power of the denominator arises from the replacement of the unit vector in equation~\ref{eq:BiotS} with the definition of a unit vector (which is $\vec{\hat{r}} = \frac{\vec{r}}{|r|}$). Note that the integral in equation~\ref{BTripInt} is over the Cartesian components of $\vec{r}$, and to be explicit $|\vec{r}|^{3} = ( \rx^{2} + \ry^{2} + \rz^{2} )^{\frac{3}{2}}$


The integral is solved using the technique of integration by parts, taking the cross product in the numerator, $\vec{J} \vec{ \times } \vec{r}$, as $\dif  g ( \rx , \ry , \rz )$ and the denominator, $\frac{1}{( \rx^{2} + \ry^{2} + \rz^{2} )^{\frac{3}{2}}}$ as $f(\rx,\ry,\rz)$ 


As we know the form of $\vec{r}$, to solve equation~\ref{BTripInt} we will need to know the volume current density vector, $\vec{J}$, it is given by $\vec{J} = \rho_v \vec{s}_{v}$, where $\rho_v$ is the charge density of oxygen vacancies and $\vec{s}_{v}$ is their average drift velocity. The charge density can be expressed as $\rho_v = \frac{n z_v}{vol}$, where $z_v$ is the charge on a single oxygen vacancy ($+1$ in this scheme because we are dealing with single oxygen vacancy in TiO$_2$ material, not the equivalent motion of an oxygen ion O$^{-}$ or the effective charge on an oxygen atom O$^{-2}$), $n_{v}$ is the number of oxygen vacancies and $vol$ is the volume. We substitute this for $\rho_v$ and substitute for velocity of the ions using standard definitions. Thus, the volume current density for all the oxygen vacancies is

\begin{equation}
\vec{J} = \frac{n_{v} z_v \mu_v \vec{L}}{vol} \;.
\label{eq:J1}
\end{equation}

Note that $L$ and $\mu_v$ are average properties, so we are dealing with the bulk movement of vacancies: individual vacancies can move at different speeds and in different directions, but drift along the field lines on average. The total charge due to the oxygen vacancies (and also our memory property), $q_v$, is $q_v = n_{v} z_v$ and so our final equation for $\vec{J}$ is 

\begin{equation}
\vec{J} = \frac{q_v \mu_v \vec{L}}{vol} \;,
\label{eq:J}
\end{equation}
and is function of time because $q_v(t)$. Note that $\vec{L}$ can also vary with time in some experiments. The vector $\vec{J}$ is taken as being
\begin{equation}
 \vec{J} = \{ \frac{q_v \mu_v L}{vol} \vec{\hat{i}} ,0 \vec{\hat{j}} , 0\vec{\hat{k}}\} \;, 
\label{eq:JVec}
\end{equation}
because field and drift direction are taken as being in the $+x$ direction for the Strukov device.

If we put equation~\ref{eq:J} into equation~\ref{BTripInt} and solve as described above we get

\begin{equation}
\vec{B}(p) = \frac{\mu_0}{4 \pi} L \mu_{v} q_v \{ P_{x}, - x z P_{y}, x y P_{z} \}
\label{eq:B}
\end{equation}

with 

\begin{eqnarray}
P_{x} & = & 0 \;, \\
P_{y} & = & \: \frac{ F}{2 \left( w^{2} + E^{2} + F^{2} \right)^{\frac{3}{2}}} \notag \\
 & & - \frac{1}{2 w E F} \frac{
a_y}
{\left( \left( w^2 + F^2 \right) b \right)} \notag \\
 & & + F \arctan \left( \frac{ w E }{F \sqrt{w^2 + E^2 + F^2} } \right) , 
\end{eqnarray}

and

\begin{eqnarray}
P_{z} & = & \: \frac{E}{2 \left( w^{2} + E^{2} + F^{2} \right)^{\frac{3}{2}}} \notag \\
 & & - \frac{1}{2 w E F} \frac{
a_z}
{\left( \left( w^2 + E^2 \right) b \right)} \notag \\
 & & + E \arctan \left( \frac{ w F }{E \sqrt{w^2 + E^2 + F^2} } \right) , 
\end{eqnarray}
where
\begin{eqnarray}
a_y & = & w E ( F^{2} \left( E^{2} + F^{2} \right)^{2} + w^{4} \left( 2 E^{2} + F^{2} \right) \notag \\
 & & + w^{2} \left( 2 E^{4} + 5 E^{2} F^{2} + 2 F^{4} ) \right) \; , \\
a_z & = & w F ( E^{2} \left( E^{2} + F^{2} \right)^{2} +
 w^{4} \left( E^{2} + 2 F^{2} \right) \notag \\
 & & + w^{2} \left( 2 E^{4} + 5 E^{2} F^{2} + 2 F^{4} ) \right) \\
b & = &  \left(E^2 + F^2 \right) \left( w^2 + E^2 + \: F^2 \right)^{\frac{3}{2}} \;.
\end{eqnarray}

$P_y$ and $P_z$ contain only the dimensions of the memristor, so even if they are not analytically simple, they are easy to calculate numerically. As expected of a magnetic field, the divergence of the field is zero, i.e. $\nabla \cdot \vec{B}=0$. 

As an example, for a Strukov memristor which is close to being full with the maximum number of vacancies (i.e. the limit) the field at point $p$ is given by

\begin{equation}
\vec{B}(p) = \{ 0, -6.37 q_v V x z, 6.37 q_v V x y \} ,
\end{equation}

where $V$ is the applied voltage, $p = \{ x, y, z \}$ and $x$, $y$ and $z$ refer to a second set of coordinates which are located outside the memristor whose unit vectors are $\Xx$, $\Yy$ and $\Zz$.
 The curl of $\vec{B}$ is non-zero as the field 
curls around the current in an anti-clockwise direction. An example curl for the system above evaluated at $\{ 0,0,0 \}$ (just inside the left hand side of the device) is $\nabla \times \vec{B}=\{ 12.74 q_v V, -6.37 q_v V, -6.37 q_v V\}$. The gradient of the field indicates the direction of travel that gives maximal field values, i.e. $\nabla \vec{B} = \{ 0, -6.37 q_v V x z, 6.37 q_v V x y \}$, namely that there is no increase in the $x$ direction and that the maximal vector field is experience by looping around the $x$ axis.

\subsection{Calculating the Magnetic Flux due to Oxygen Vacancy Current}

The magnetic $\vec{B}$ field is the magnetic flux density and so to calculate the magnetic flux through a surface associated with this field, $\varphi$, we need to take the surface integral

\begin{equation}
\varphi = \int \vec{B} \cdotp \dif \vec{A}
\label{eq:B.dA}
\end{equation}

where $\dif \vec{A}$ is the normal vector from the surface infinitesimal $\dif A$. 

As it is a surface integral, to calculate the magnetic flux we need to pick a surface to evaluate over. It makes sense to choose a surface that correlates to one of the surfaces of the device. Picking the surface just above the device ($0<x<D$, $0<y<E$, $z=F$), we use the surface normal area infinitesimal, $\vec{\dif A}$, which is given by $\vec{\dif A} = \{0,0,\hat{\imath} \hat{\jmath}\}$. As is standard in electromagnetism, we integrate over the entire area. The limits of the surface are taken to be the dimensions of the device.

By putting the expression for $\vec{B}$ in equation~\ref{eq:B} into equation~\ref{eq:B.dA} and taking the surface integral, we derive the general form of the magnetic flux passing through a surface $i$-$j$:

\begin{equation}
	\varphi=\frac{\mu_{0}}{4 \pi} L \mu_{v} i j P_{k} q_{v} \:  ,
	\label{eq:1}
\end{equation}	
where $i \epsilon \{ x, y, z \}$, $j \epsilon \{ x, y, z \}$, $k \epsilon \{ x, y, z \}$, i.e. $P_k$ is component in the vector in equation~\ref{eq:B} which is perpendicular to the surface $i-j$ where $i$ and $j$ can be any two of the Cartesian directions.

Equation~\ref{eq:1} contains a physical magnetic flux, satisfies Chua's equation $\varphi=M(q)q$~\cite{14} and crucially has been derived without reference to Chua's equations. Note that this relation between charge and flux in a memristor includes the material properties and is the first to do so. 

By reference to equation~\ref{eqn:Chua}, the Chua memristance in this device is expressed as:

\begin{equation}
	M \left( q_{v} \left( t \right) \right) =  U X \mu_{v} P_{k}\left( q_{v} \left( t \right) \right) \: , 
\label{eq:2}
\end{equation} 
where we have gathered up the constants and explicitly included $P_k$'s dependence on $q_v$.

Equation~\ref{eq:2} can be considered as three separate parts: 

\begin{enumerate}
 \item $U$, the universal constants: $\frac{\mu_0}{4 \pi}$, this term includes the effects of the permittivity of a vacuum on memristance. It's inclusion in the equation clearly demonstrates that magnetism is involved in memristance.
 \item $X$, the experimental constants: $D E L$, where $D E$ is the surface the flux was calculated over as we've substituted in for $i$ and $j$, in this case the top of the device. The constant $X$ will be different for different devices and experiments and is time-dependent if $V$ is. 
 \item $\beta$, the material variable: $\mu_{v} P_{z}$, this includes the physical dimensions of the device, but it will change throughout the experiment as a result of the moving boundary, $w(t)$, whose motion is caused by the drift of vacancies across the device. This is the only term that contains variables. Note, it is from this term, via the value of $\mu_v$ and its interaction with the applied voltage frequency that the memristor's frequency dependence arises. 
\end{enumerate}

For the Strukov memristor, $P_{y}$ and $P_{z}$ are equal in magnitude because the magnetic field is centro-symmetric around the vacancy current (which flows in the $+x$ direction, see figure~\ref{fig1}). Thus, the values of the memristance calculated from the $x$-$y$ and $x$-$z$ surfaces are the same, see Table~\ref{tab:SurfacePhi}. As $w$ is a measure of how far the vacancies extend into the material it is dependent on $q_v$ and thus $P_{k}$ is a function of $q_v$. Interestingly, equation~\ref{eq:2} implies that the Chua memristance has directional dependence, and will only be non-zero for surfaces that cut magnetic field lines, the $y$-$z$ surface doesn't, and thus $P_x$ is zero. This raises the intriguing possibility of memristance being best described as a three-dimensional property. For most systems there will be only one non-zero value. As Chua suggested that the memristance could be either charge or flux controlled~\cite{84}, the memristance calculated here should be capable of being controlled by 
either 
and thus holding a memory of either. And it does, $P_{k}(q_v)$ is part of the Chua memristance which holds the memory of the charge, $\beta$, is part of the memristance which holds the memory for the flux.

\begin{table}[!t]
\renewcommand{\arraystretch}{1.5}
\begin{tabular}{|c|c|l|c|}
\hline
 Device 	& Area 	& Integral 	& Value	\\
 surface 	& infinitesimal	&  	& for Strukov	\\
 	& $\hat{\dif \vec{A}} $	& 		& memristor	\\
\hline
 Top		& $\{ 0, 0, \Xx \Yy \}$ & $\varphi_{\mathrm{top}} = $ & -3.186$\times 10^{-15} q_v$ \\
		&  &   $\int_{0}^{E} \int_{0}^{D} \vec{B} \cdot \hat{\dif \vec{A}} \dif x \dif y$ &  \\
\hline
 Bottom		& $\{ 0, 0, - \Xx \Yy \}$ & 
$\varphi_{\mathrm{bottom}} =$ &
-3.186$\times 10^{-15} q_v$ \\
 & & $\int_{0}^{E} \int_{0}^{D} \vec{B} \cdot \hat{\dif \vec{A}} \dif x \dif y$ & \\
\hline
 Front		& $\{ 0,  \Xx \Zz, 0 \}$ & 
$\varphi_{\mathrm{front}} = $ &
-3.186$\times 10^{-15} q_v$ \\
		&  & 
 $\int_{0}^{F} \int_{0}^{D} \vec{B} \cdot \hat{\dif \vec{A}} \dif x \dif z$ & \\
\hline
 Back		& $\{ 0,  - \Xx \Zz, 0 \}$ & 
$\varphi_{\mathrm{back}} = $&
3.186$\times 10^{-15} q_v$ \\
		& & $\int_{0}^{F} \int_{0}^{D} \vec{B} \cdot \hat{\dif \vec{A}} \dif x \dif z$ & \\
\hline
 Left		& $\{ \Yy \Zz, 0, 0 \}$ & 
$\varphi_{\mathrm{left}} = $ &
0 \\
		&  & $\int_{0}^{F} \int_{0}^{E} \vec{B} \cdot \hat{\dif \vec{A}} \dif y \dif z$ & \\
\hline
 Right		& $\{ -\Yy \Zz, 0, 0 \}$ & 
$\varphi_{\mathrm{right}} = $ &
0 \\
	&  & $\int_{0}^{F} \int_{0}^{E} \vec{B} \cdot\hat{\dif \vec{A}} \dif y \dif z$ &
 \\
\hline
\end{tabular}
\caption{Table for the magnetic flux as calculated from the different possible surfaces of the memristor.}
\label{tab:SurfacePhi}
\end{table}

Putting in real-world values for the device characteristics (as above, including $V=1V$) for the Strukov memristor gives a memristance equation of $d \varphi = 3.53 \times10^{15} dq$, and a $\varphi - q$ plot is linear over the range $0< w < D$ (where $w$ must be strictly more than 0 to avoid 1/0 errors), indicating this model is a perfect memristor because it satisfies Chua's constitutive definition (equation~\ref{eqn:Chua}) with a constant value.

With these real-world example values, the Stukov memristors has flux of 2.44$\times 10^{-29}$Wb. In contrast, the magnetic flux associated with the conducting electrons through the same surface is -4.07$\times 10^{-24}$Wb. This is in the opposite direction and approximately 100 000 times bigger than the vacancies' magnetic flux. This may explain why the magnetic flux associated with memristor function has not been experimentally measured.


\section{Memory and Conservation Functions}

How can the tiny ($\sim 10^{-29}$Wb) magnetic flux in the Strukov memristor be associated with the large effect seen in experimental I-V curves? The answer is because the memristive movement of charge affects the resistivity of the material, and it is this resistivity change that is `sampled' by the conducting electronic current. 

When measuring a memristor it is conventional to measure the electronic current, not the ionic current. As the electronic current is many times larger and faster than the movement of vacancies, we can even choose to ignore the vacancy contribution to the total flow of charge, without introducing a significant error. What is needed is the memristance as experienced by the conducting electrons, $R_{\mathrm{tot}}(t)$. The component of that memristance which is directly due to the changing resistivity of the doped material, $M_{e}$, which we shall call the Memory Function as it encapsulates the memristor's memory, is given by

\begin{equation}
M_{e} = C  M(q_{v}(t)) \: ,
\end{equation}

where $C$ is an experimentally determined parameter for the material.

Because the ion mobilities of the electrons, $\mu_{e}$, and the vacancies, $\mu_{v}$, are measured experimentally, it is predicted that 
$C = \left( q_{e} \mu_{e} \right) / \left( q_{v} \mu_{v} \right)$.
	
The memory function describes the doped part of the titanium dioxide, TiO$_{(2-x)}$, as experienced by the electrons traversing it. The proportion of the memristor made up of this form changes, and, because matter must be conserved in the model, the proportion of the memristor made up of un-doped TiO$_{2}$ is given by the conservation function, $R_{\mathrm{con}}$, which is simply the resistance of the un-doped material:

\begin{equation}
	R_{\mathrm{con}} \left( t \right) = \frac{\left( D - w \left( t \right) \right) \rho_{TiO_{2}}}{E F}	
	\label{eq:5}
\end{equation}

where $\rho_{TiO_{2}}$ is the resistivity of un-doped TiO$_2$. Note, Strukov et al's model was based on a similar conservation function (as it arises from Ohm's law) and, as this is responsible for most of the observed change in the device, their model gave memristor $I-V$ curves.

The total resistance as experienced by the conducting electrons, $R_{\mathrm{tot}}$, is then given by
\begin{equation}
  R_{\mathrm{tot}} = R_{\mathrm{con}} + M_{e} \; .
\end{equation}

As $R_{\mathrm{tot}}$ is a resistance that changes with time due to the action of charge it is therefore also a memristance and this equation gives the pinched hysteresis loop in $I$-$V$ space which is indicative of memristance, as shown in figure~\ref{fig:2}. Separately, both the conservation and memory functions are also memristances and both can give rise to a memristive $I$-$V$ curve. The memory function is just the Chua memristance expressed in terms of the conducting electrons. The conservation function is memristance due to the change in volume of the un-doped material, which is the second effect of the oxygen vacancies' movement into the TiO$_2$. 

A Chua memristor is that it is a function of a single state variable~\cite{84} (compared to a memristive system, which can have more than one). The only variable in the conservation function is $w$ and because $w$ is a measure of how far the memristive charges have moved, the Chua memristance, and thus the memory function, can be written in terms of $w$ instead of $q$. Therefore, $R_{\mathrm{tot}}$ can be written as a function of $w$ only, thus demonstrating that the Strukov memristor is a Chua memristor with one state variable $w$. Assuming that the vacancies are spread out in the same way across the device, (ie that TiO$_{(2-x)}$) $w$ is a measure of $q$ and these equations can also be expressed in terms of a single state variable $q$.
	
\begin{figure}[!t]
\includegraphics[width=3.5in]{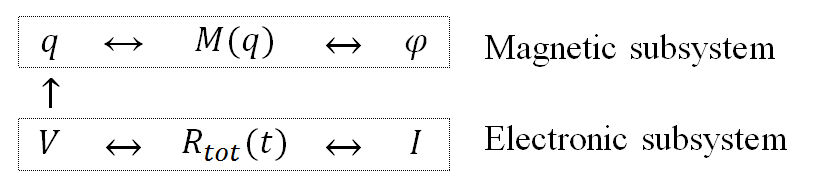}
\caption{\label{fig:3} Diagrammatic representation of the presented memristor model. The Magnetic subsystem refers to the magnetic flux required by Chua's definition of memristance~\cite{14} and the non-electronic charge carriers that give rise to it. The Electronic subsystem refers to the effects experienced by the conducting electrons. Shown here is an example of a voltage controlled memristor where the applied voltage causes the non-electronic carriers to drift and their movement effects local resistivity of the material, altering the total resistance $R_{\mathrm{tot}}$ and affecting the measured electronic current.}
\end{figure}

\begin{figure}[!t]
\includegraphics[width=3.2in]{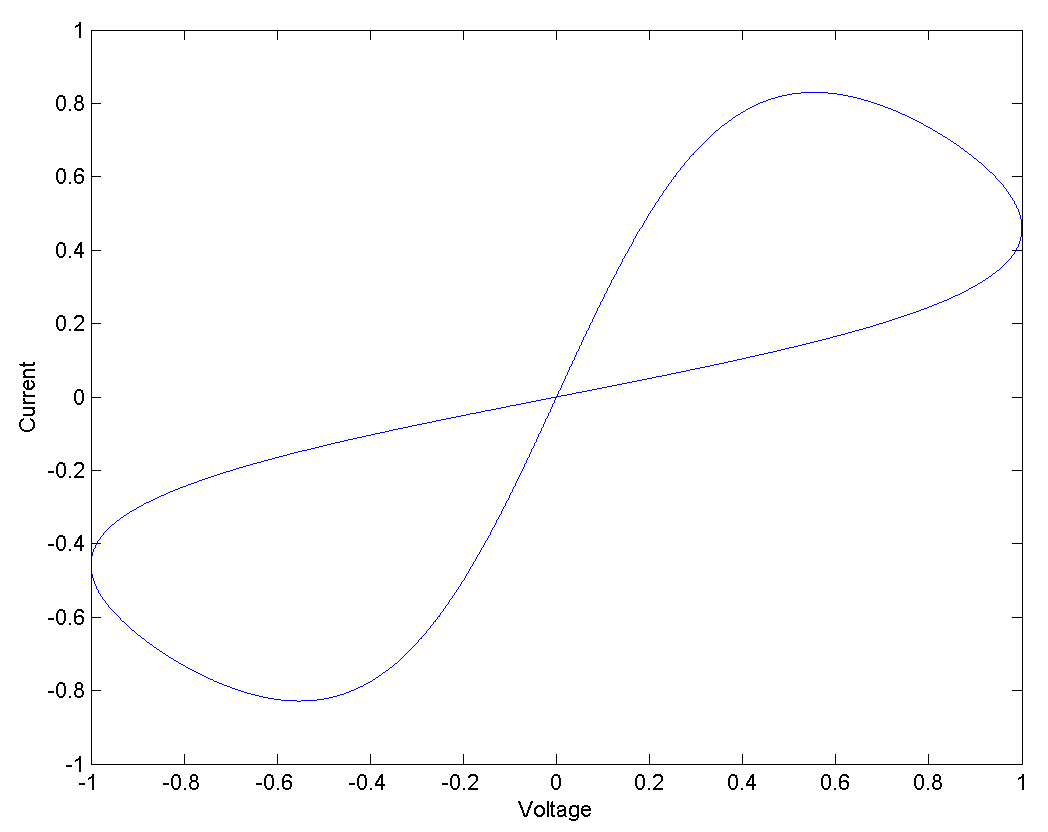}
\caption{\label{fig:2} An example memristor $I$-$V$ curve as calculated from the presented theory.}
\end{figure}

Thus we have demonstrated that in order to describe memristance, two systems need to be considered, as is shown diagrammatically in figure~\ref{fig:3}. The first is the `electronic' system, which is associated with the conducting electrons and which provides the `electronic current' response to an applied voltage. 

The second system is the `magnetic' system, which contains the magnetic flux and the `memristive' charge, i.e. the vacancies. Note that these charge carriers are not especially magnetic (neither is it claimed here that the memristive charge carriers are acting as magnetic monopoles, although that comparison has been made~\cite{118}). Instead, the charge responsible for the memory function of the memristor is being separated conceptually from the charge due to the conducting electrons. It is important to realise that the existence of memristive magnetic flux in the system does not mean that the memristor is magnetised in a traditional sense. The `magnetism' in the system is not similar to the magnetism of ferrous materials that are capable of holding or reacting strongly to a magnetic field. Instead the memristor magnetic effect is similar to the atomic scale magnetic susceptibility as understood and exploited by NMR spectroscopy and MRI imaging. Furthermore, the `magnetic' system does not describe all of 
the properties of the memristor that exhibit magnetism. For example, there is magnetic flux associated with the conducting electrons, but this flux is mostly irrelevant to understanding the memristive operation of the device. 

\section{Conclusion}
Thus, we have shown that by modelling the magnetic flux associated with the oxygen vacancies in titanium dioxide, we can satisfy Chua's constitutive equation for the memristor. This new model has some advantages over previous models. It includes the ions in a natural way, solves the question of the missing magnetic flux, models the entire device and links device properties directly to Chua's memristor equations -- which might aid future device design. The drawback of this model is that it is more complex than previous models, as it requires two conducting species to be considered and the integration procedure, although relatively straightforward once split into parts, is not as obvious or easy to solve as other models. However, the model presented in this paper has been experimentally verified~\cite{G1}.

This work does present an intriguing idea which I am not certain has been considered at great length in the literature. We convert from the resistance as felt by the ions to that felt by the electrons, implicitly considering resistance as a phenomenon which is dependent on the charge carrying species under consideration. This idea appears in chromatography and ion mobility spectroscopy, but seems to be neglected within modern electronic engineering.


\section*{Acknowledgment}
The author would like to thank Leon Chua for helpful discussions, Ben de Lacy Costello, Oliver Matthews and Deborah Gater for critiques. EPSRC funded this work.


\def\IEEEbibitemsep{1pt plus .5pt}

\bibliographystyle{IEEEtran}
\bibliography{IEEEabrv,UWELit}

\begin{thebibliography}{10}
\providecommand{\url}[1]{#1}
\csname url@samestyle\endcsname
\providecommand{\newblock}{\relax}
\providecommand{\bibinfo}[2]{#2}
\providecommand{\BIBentrySTDinterwordspacing}{\spaceskip=0pt\relax}
\providecommand{\BIBentryALTinterwordstretchfactor}{4}
\providecommand{\BIBentryALTinterwordspacing}{\spaceskip=\fontdimen2\font plus
\BIBentryALTinterwordstretchfactor\fontdimen3\font minus
  \fontdimen4\font\relax}
\providecommand{\BIBforeignlanguage}[2]{{%
\expandafter\ifx\csname l@#1\endcsname\relax
\typeout{** WARNING: IEEEtran.bst: No hyphenation pattern has been}%
\typeout{** loaded for the language `#1'. Using the pattern for}%
\typeout{** the default language instead.}%
\else
\language=\csname l@#1\endcsname
\fi
#2}}
\providecommand{\BIBdecl}{\relax}
\BIBdecl

\bibitem{278}
M.~Kumar, ``Memristor - why do we have to know about it?'' \emph{IETE Tech.
  Rev.}, vol.~26, pp. 3--6, 2009.

\bibitem{84}
L.~O. Chua and S.~M. Kang, ``Memristive devices and systems,''
  \emph{Proceedings of the IEEE}, vol.~64, pp. 209--223, 1976.

\bibitem{247}
L.~Chua, V.~Sbitnev, and H.~Kim, ``Hodgkin-huxley axon is made of memristors,''
  \emph{International Journal of Bifurcation and Chaos}, vol.~22, p. 1230011
  (48pp), 2012.

\bibitem{248}
------, ``Neurons are poised near the edge of chaos,'' \emph{International
  Journal of Bifurcation and Chaos}, vol.~11, p. 1250098 (49pp), 2012.

\bibitem{51}
A.~Smerieri, T.~Berzina, V.~Erokhin, and M.~P. Fontana, ``Polymeric
  electrochemical element for adaptive networks: Pulse mode,'' \emph{Journal of
  Applied Physics}, vol. 104, p. 114513, 2008.

\bibitem{3}
Y.~V. Pershin, S.~L. Fontaine, and M.~di~Ventra, ``Memristive model of amoeba's
  learning,'' \emph{Phys. Rev. E}, vol.~80, p. 021926 (6 pages), 2009.

\bibitem{239}
C.~Zamarreno-Ramos, L.~A.~C. {n}as, J.~A. P\'{e}rez-Carrasco, T.~Masquelier,
  T.~Serrano-Gotarredona, and B.~Linares-Barranco, ``On spike-timing dependent
  plasticity, memristive devices and building a self-learning visual cortex,''
  \emph{Frontiers in Neuormorphic engineering}, vol.~5, pp. 26(1)--26(20),
  2011.

\bibitem{DavidJ1}
G.~D. Howard, E.~Gale, L.~Bull, B.~de~Lacy~Costello, and A.~Adamatzky,
  ``Evolution of plastic learning in spiking networks via memristive
  connections,'' \emph{IEEE Transactions on Evolutionary Computation}, vol.~16,
  pp. 711--719, 2012.

\bibitem{14}
L.~O. Chua, ``Memristor - the missing circuit element,'' \emph{IEEE Trans.
  Circuit Theory}, vol.~18, pp. 507--519, 1971.

\bibitem{279}
S.~P. Adhikari, H.~K. Maheshwar Pd.~Sah, and L.~O. Chua, ``Three fingerprints
  of memristor,'' \emph{IEEE Transactions on Circuits and Systems}, vol.~60,
  pp. 3008--3021, November 2013.

\bibitem{249}
L.~Chua, ``The origin of complexity.''

\bibitem{93}
Y.~V. Pershin and M.~D. Ventra, ``Memory effects in complex materials and
  nanoscale systems,'' \emph{Advances in Physics}, vol.~60, pp. 145--227, 2011.

\bibitem{183}
F.~Argall, ``Switching phenomena in titanium oxide thin films,'' \emph{Solid
  State Electronics}, vol.~11, pp. 535--541, 1968.

\bibitem{12}
V.~Erokhin, T.~Berzina, and M.~P. Fontana, ``Hybrid electronic device based on
  polyaniline-polyethylenoxide junction,'' \emph{J. Appl. Phys.}, vol.~97, p.
  064501, 2005.

\bibitem{5}
V.~Erokhin and M.~Fontana, ``Electrochemically controlled polymeric device: a
  memristor (and more) found two years ago,'' \emph{arXiv:0807.0333v1
  [cond-mat.soft]}, 2008.

\bibitem{119}
L.~Chua, ``Resistance switching memories are memristors,'' \emph{Applied
  Physics A: Materials Science \& Processing}, pp. 765--782, 2011.

\bibitem{236}
C.~T. Themistoklis~Prodromakis and L.~Chua, ``Two centuries of memristors,''
  \emph{Nature Materials}, vol.~11, pp. 478--481, 2012.

\bibitem{280}
J.-M. Ginoux and B.~Rossetto, \emph{The Singing Arc: The Oldest Memristor?},
  A.~Adamatzky and G.~Chen, Eds.\hskip 1em plus 0.5em minus 0.4em\relax World
  Scientific, 2012.

\bibitem{15}
D.~B. Strukov, G.~S. Snider, D.~R. Stewart, and R.~S. Williams, ``The missing
  memristor found,'' \emph{Nature}, vol. 453, pp. 80--83, 2008.

\bibitem{142}
\BIBentryALTinterwordspacing
\emph{The mythology of the memristor,}.\hskip 1em plus 0.5em minus 0.4em\relax
  ISCAS, 2010. [Online]. Available:
  \url{http://www.slideshare.net/blaisemouttet/mythical-memristor}
\BIBentrySTDinterwordspacing

\bibitem{71}
S.~H. Jo, T.~Chang, I.~Ebong, B.~B. Bhadviya, P.~Mazumder, and W.~Lu,
  ``Nanoscale memristor device as a synapse in neuromorphic systems,''
  \emph{Nanoletters}, vol.~10, pp. 1297--1301, 2010.

\bibitem{255}
E.~M. Gale, B.~{de Lacy Costello}, and A.~Adamatzky, ``The effect of electrode
  size on memristor properties: An experimental and theoretical study,'' in
  \emph{2012 IEEE International Conference on Electronics Design, Systems and
  Applications (ICEDSA 2012)}, Kuala Lumpur, Malaysia, Nov. 2012.

\bibitem{M0}
E.~Gale, R.~Mayne, A.~Adamatzky, and B.~de~Lacy~Costello, ``Drop-coated
  titanium dioxide memristors,'' \emph{Materials Chemistry and Physics}, vol.
  143, pp. 524--529, January 2014.

\bibitem{29}
B.~Widrow, ``An adaptive 'adaline' neuron using chemical 'memistors',''
  \emph{Technical Report}, 1960.

\bibitem{124}
B.~O. Aduda, P.~Ravirajan, K.~L. Choy, and J.~Nelson, ``Effect of morphology on
  electron drift mobility in porous tio$_2$,'' \emph{International Journal of
  Photoenergy}, 2004.

\bibitem{145}
Y.~V. Pershin and M.~D. Ventra, ``Spin memristive systems: Spin memory effects
  in semiconductor spintronics,'' \emph{Phys. Rev. B}, vol.~78, pp. 113\,309--1
  -- 113\,309--4, 2008.

\bibitem{144}
T.~Driscoll, H.-T. Kim, B.~G. Chae, M.~D. Ventra, and D.~Basov,
  ``Phase-transition driven memristive system,'' \emph{Appl. Phys. Lett.},
  vol.~95, p. 043503, 2009.

\bibitem{Electrodynamics}
D.~Griffiths, \emph{Introduction to Electrodynamics}, third edition,
  international edition,~ed.\hskip 1em plus 0.5em minus 0.4em\relax San
  Francisco: Pearson Benjamin Cummings, 2008.

\bibitem{118}
Y.~Z. Umul, ``Magnetic charges that behave as magnetic monopoles,''
  \emph{Progress in Electromagnetics Research Letters}, vol.~19, pp. 19--28,
  2010.

\bibitem{G1}
B.~d. L.~C. Ella~Gale and A.~Adamatzky, ``Which memristor theory is best for
  relating device properties to memristive function: A test of three
  theories,'' \emph{Submitted}.

\bibitem{M1}
E.~Gale, D.~Pearson, S.~Kitson, A.~Adamatzky, and B.~de~Lacy~Costello, ``The
  effect of changing electrode metal on solution-processed flexible titanium
  dioxide memristors,'' \emph{Submitted}, 2013.

\bibitem{45}
V.~Erokhin, A.~Schuz, and M.~Fontana, ``Organic memristor and bio-inspired
  information processing,'' \emph{International Journal of Unconventional
  Computing}, vol.~6, pp. 15--32, 2009.

\bibitem{81}
A.~Smerieri, V.~Erokhin, and M.~P. Fontana, ``Origin of current oscillations in
  a polymeric electrochemically controlled element,'' \emph{Journal of Applied
  Physics}, vol. 103, p. 094517, 2008.

\bibitem{6}
V.~Erokhin, T.~Berzina, P.~Camorani, and M.~P. Fontana, ``Conducting polymer -
  solid electrolyte fibrillar composite material for adaptive networks,''
  \emph{Soft Matter}, vol.~2, pp. 870--874, 2006.

\bibitem{254}
E.~M. Gale, B.~{de Lacy Costello}, and A.~Adamatzky, ``Filamentary extension of
  the {Mem-Con} theory of memristance and its application to titanium dioxide
  {Sol-Gel} memristors,'' in \emph{2012 IEEE International Conference on
  Electronics Design, Systems and Applications (ICEDSA 2012)}, Kuala Lumpur,
  Malaysia, Nov. 2012.

\end{thebibliography}
%





\end{document}